\documentclass[prl,a4paper,twocolumn,showpacs,amssymb]{revtex4}
\usepackage{graphicx}

\begin {document} 
%\draft %Obsolite in revtex4

\title{Note on the superfluid Reynolds number for turbulent flow of superfluid $^4$He around an oscillating sphere}

\author{W. Schoepe}

\affiliation{Fakult\"at f\"ur Physik, Universit\"at Regensburg, D-93040 Regensburg, Germany}

\begin{abstract}
The superfluid Reynolds number $Re_s = (v - v_c)\,D/\kappa$ can be expressed simply by the number of vortex rings that are shed during a half-period of the oscillation.

\end{abstract}


\maketitle %New revtex4 format

In a recent work by Reeves et al.\,\cite{Reeves} the two-dimensional Gross-Pitaevskii equation was investigated numerically in the vicinity of the critical velocity $v_c$ for the onset of quantum turbulence. The central result of that work was the observation of a dynamic similarity in the wake of a cylindrical object and its breakdown due to vortex shedding at a superfluid Reynolds number is given by
\begin{equation}
Re_s = \frac{(v-v_c)\,D}{\kappa},
\end{equation}
where $v_c$ is the critical velocity for the onset of turbulence, $D$ is a characteristic length scale, and $\kappa$ is the circulation quantum. Applying this result to our experiments with a sphere oscillating in $^4$He below 0.5\,K (for a recent review, see \cite{review}), we choose $D = 2\,R$ as the characteristic length scale, where $R$ = 0.12 mm is the radius of the sphere. Defining $\Delta v \equiv (v-v_c)$ and $v_0 \equiv \kappa /2\,R$ we write
\begin{equation}
Re_s = \frac{\Delta v}{v_0},
\end{equation}
where in our case $v_0 = 0.40$ mm/s. 
This result is valid for a sphere, but no assumptions have been made concerning the dimension of the flow (2D or 3D)  nor of its type (steady or oscillatory). We note, that the ratio $\kappa /R$ determines the self-induced velocity of a vortex ring of radius $R$.

In the following we show from the data analysis of our experiments, how Eq.(2) can be interpreted in very simple way, namely that $Re_s$ is given by the number $n$ of vortex rings that are shed from the sphere during one-half period of the oscillation $1/(2f)$, where $f$ = 119 or 160 Hz is the frequency of the oscillating sphere.

In a small interval of $\Delta v $ above $v_c \approx $ 20 mm/s with $\Delta v/v_c \le$ 0.03, we
observe that the flow pattern is unstable, switching intermittently between turbulent phases and potential flow. These patterns are easily identified because the drag force on the sphere is much larger in the turbulent regime than during potential flow. Recording time series at constant temperature and driving force $F$, we analyze the distribution of the lifetimes $t$ of the turbulent phases and find an exponential distribution $\exp(-t/\tau)$, and mean lifetimes $\tau$ increasing very fast with the driving force amplitude, namely as
\begin{equation}
\tau = \tau_0\,\exp[\,(F/F_1)^2],
\end{equation}
\noindent where the fitting parameters $\tau_0$ = 0.5 s at 119 Hz and 0.25 s at 160 Hz, and $F_1$ = 18 pN and 20 pN, respectively. The force $F_1$ can interpreted as being caused by the loss of kinetic energy of the sphere due to the shedding of {\it one} vortex ring of radius $R$ during {\it one} half-period.\cite{34,review} From a fit to the data we find
\begin{equation}
F_1 = 1.3 \rho \kappa R \sqrt{\kappa \omega}, 
\end{equation}
\noindent where $\rho$ is the density of the liquid and $\omega = 2\pi f$. The driving force is obtained from the data $v(F)$ and is given by 
\begin{equation}
F(v) = (8/3\pi)\gamma (v^2 - v_c^2).
\end{equation}
\noindent $\gamma$ is identical to the expression for classical turbulent flow around a sphere, namely $\gamma = c_D\rho \pi R^2/2$ and the drag coefficient of a sphere is $c_D \approx $ 0.4.\cite{Landau} The numerical factor $8/3\pi$ = 0.85 takes into account the energy balance for an equilibrium oscillation amplitude: energy gain from the drive and loss from a quadratic damping must cancel. While Eq.(5) is deduced from the experiment up to velocities of ca.\,100 mm/s, which is 5 times larger than $v_c$, Eq.(4) is proven valid only in the small interval $\Delta v/v_c \le$ 0.03 where $\tau $ was measurable. In this regime we may approximate Eq.(5) by  
\begin{equation}
F(v) = (8/3\pi)\,2\gamma \, v_c\,\Delta v.
\end{equation}
\noindent We assume that the number $n = F/F_1$ is the average number of vortex rings emitted per half-period. Inserting Eq.(4) and Eq.(6), and using our results $v_c = 2.8 \sqrt{\kappa \omega}$ , we find
\begin{equation}
n = \frac{F}{F_1} = \frac{(8/3\pi)\,2 \gamma  v_c \, \Delta v}{1.3\,\,\rho \,\kappa \,R\, \sqrt{\kappa \,\omega}} = \frac{\Delta v}{v_1}, 
\end{equation}
where $v_1 = 0.48\,\kappa /R$ = 0.39 mm/s.\\ 
\noindent In Fig.1 we plot the normalized mean lifetime 
\begin{equation}
\tau^*(\Delta v) \equiv \tau /\tau_0 = \exp{[(\Delta v/v_1)^2]}. 
\end{equation}
The salient feature is that $\tau^*$ is independent of the oscillation frequency, of the temperature, and is not affected by $^3$He impurities. The only frequency dependence is in $\tau_0$.

\begin{figure}[t]
\includegraphics[width=1.15\linewidth]{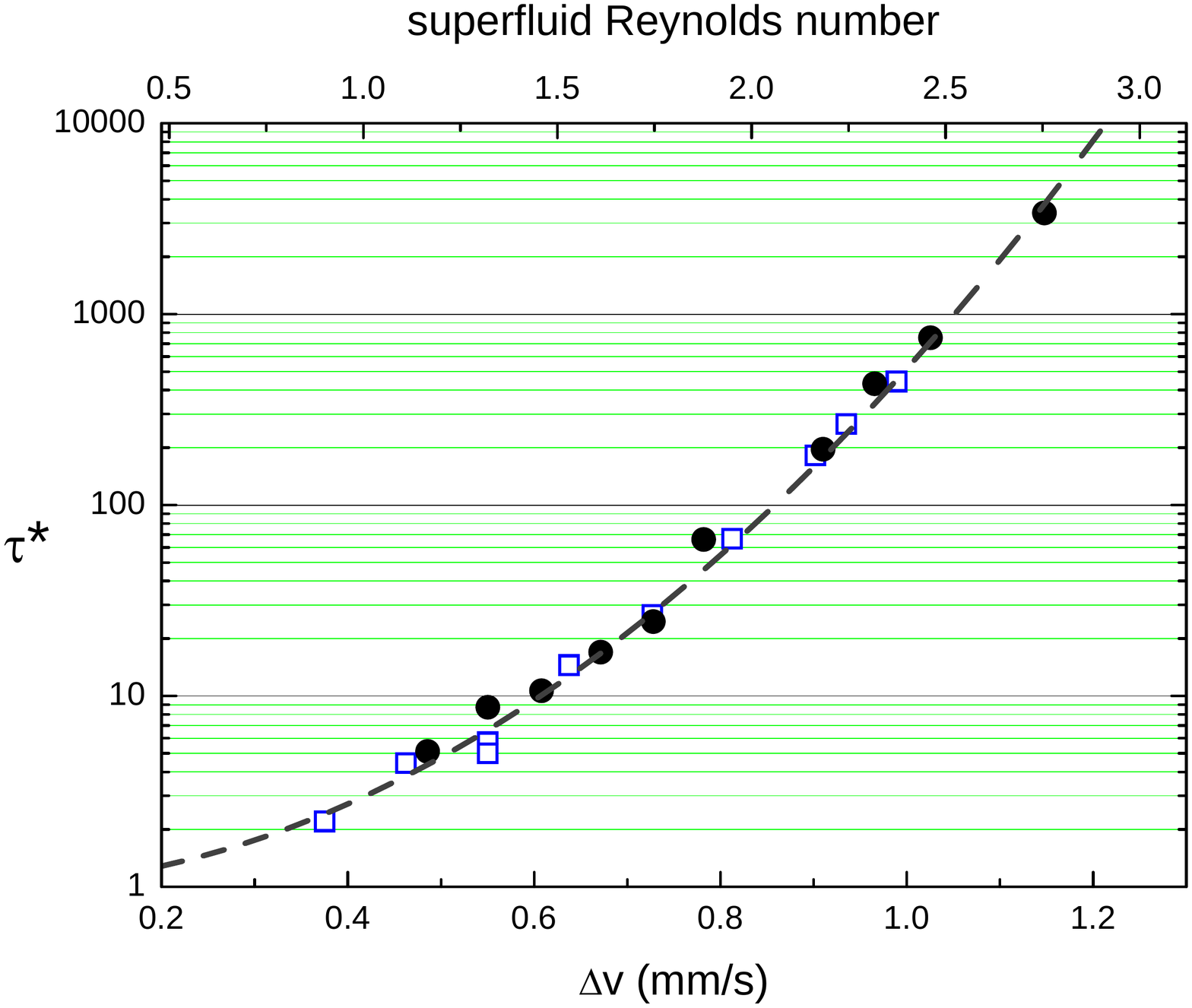}
%\centerline{\includegraphics[width=1.3\columnwidth,clip=true]{Graph4correct.pdf}}
%\includegraphics[width=1.0\textwidth]{figure1.eps}
% figure caption is below the figure
\caption{(Color online, from \cite{review}) The normalized mean lifetimes $\tau ^* = \tau /\tau_0$ as a function of $\Delta v = v - v_c$ for the 119 Hz oscillator at 301 mK (blue squares) and the 160 Hz oscillator at 30 mK with 0.05\% $^3$He (black dots). Note the rapid increase of $\tau ^*$ by 3 orders of magnitude over the small velocity interval of ca. 0.7 mm/s\,. The frequency, the temperature, and the $^3$He concentration have no effect on the data. The dashed curve is calculated from Eq.(8).}
%\label{fig:1
\end{figure}
\noindent Moreover, we see that within our estimated experimental resolution of about 10\% (from the accuracy of the numerical factors of $v_c$ and $F_1$ in Eq.(7)), the velocities $v_0$ and $v_1$ are identical. Hence, we have our main result:
\begin{equation}
Re_s = n. 
\end{equation}

That means, in our experiments (where $Re_s < $ 3, see Fig.1) the superfluid Reynolds number is given by the number of vortex rings that are shed from the sphere during one-half period of the oscillation. This is a surprisingly simple result.
Because Eq.(2) is a rather general expression it may be possible that Eq.(9) remains valid for larger values of $Re_s$ as well. But this remains to be proven.

Finally, it should be mentioned that in a completely different context an equally simple superfluid Reynolds number has recently been calculated for 2D superfluid turbulence in the limit of $Re_s\gg $ 1 to be given by the number of 2D vortices.\cite{Ash} 

\bigskip
e-mail: wilfried.schoepe@ur.de

%\end{multicols*} % ###multicol###

\end{document}